\documentclass[useAMS,usenatbib,usenatNishiNishibib,usegraphicx]{mn2e}

\newcommand{\aap}{A\&A}

\newcommand{\apjl}{ApJ}

\newcommand{\apjs}{ApJSS}

\title[]{Overlapping abundance gradients and azimuthal gradients related to the spiral structure
of the Galaxy}
 
 \author[L\'epine et al.]{ J.R.D. L\'epine$^{1}$\thanks{E-mail: jacques@astro.iag.usp.br},
  P. Cruz$^{2}$, S. Scarano Jr.$^{1}$, D.A. Barros$^{1}$, W.S.Dias$^{3}$, L. Pomp\'eia$^{4}$,\\
  \and  S.M.Andrievsky$^{5,6}$, G. Carraro$^{7}$, B. Famaey $^{8,9}$\\ \\
 $^{1}$Instituto de Astronomia, Geof\'isica e Ci\^encias Atmosf\'ericas, Universidade
 de S\~ao Paulo, Cidade Universit\'aria, S\~ao Paulo, SP, Brazil\\ 
 $^{2}$Departamento de Astrofisica, Centro de Astrobiologia (CAB/INTA-CSIC),Madrid, Spain \\
 $^{3}$UNIFEI, DFQ - Instituto de Ciˆencias Exatas, Universidade Federal de Itajub´a, Itajub´a, MG, Brazil\\
 $^{4}$Institut d'Astronomie et d'
 Astrophysique, Universit\'e Libre de Bruxelles, CP 226, Boulevard du Triomphe, 1050 Bruxelles, Belgium\\
 $^{5}$Department of Astronomy and Astronomical Observatory, Odessa National University, and Isaac Newton
 Institute of Chile \\ Odessa branch, Shevchenko Park, 65014, Odessa, Ukraine\\
 $^{6}$ GEPI, Observatoire de Paris-Meudon,  CNRS,  Universit\'e Paris Diderot,
   F-92125 Meudon Cedex, France \\
 $^{7}$ESO, Alonso de Cordova 3107, Santiago de Chile, Chile\\
 $^{8}$Observatoire Astronomique, Universit\'e de  Strasbourg, CNRS UMR 7550, 67000 Strasbourg, France \\
 $^{9}$AIfA, Universit\"at Bonn, Bonn, Germany \\}
 
\begin{document}

%\date{Accepted  Received }

%\pagerange{\pageref{firstpage}--\pageref{lastpage}} \pubyear{2009}

\maketitle

\begin{abstract}

The connection between some features of the metallicity gradient in the Galactic disk,
best revealed by  Open Clusters and  Cepheids, and the spiral structure, is explored. 
The step-like abrupt decrease in metallicity at 8.5 kpc (with $R_0$= 7.5 kpc, or at
9.5 kpc if $R_0$ = 8.5 kpc is adopted) is well explained by the corotation ring-shaped 
gap in the density of gas, which isolates the internal and external regions of the disk
one from the other. This solves a long standing problem of understanding the 
different chemical characteristics of the inner and outer parts of the disk. 
The time required to build up the metallicity difference between the two sides
of the step is a measure of the minimal life-time of the present grand-design spiral
pattern structure, of the order of 3 Gyr. The plateaux 
observed on each side of the step are interpreted in terms of the large scale 
radial motion of the stars and of the gas flow induced by the spiral structure. 
The star-formation rate revealed by the density of open clusters is maximum
in the Galactic radial range from 6 to 12 kpc (with an exception of a narrow gap
at corotation), coinciding with the region where the 4-arms mode is allowed to
exist. We argue that most of the old open clusters situated at large galactocentric
radii were born in this inner region where conditions more favorable to star-formation
are found. The ratio of $\alpha$-elements to Fe of the sample of Cepheids does not
 vary appreciably with the Galactic radius, which reveals an homogeneous history of
 star formation.  Different arguments are given showing that usual approximations
 of chemical evolution models, which assume fast mixing of metallicity in the 
 azimuthal direction and ignore the existence of the spiral arms, are a poor ones.

\end{abstract}


\section{Introduction}

The variation of the abundance of chemical elements as a function of radius in the Galactic
disk provides constraints to star formation history in the disk and to  basic 
assumptions made in chemical evolution  models. Several reviews have been published 
recently, focusing in different chemical elements like Iron or $\alpha$-elements, and in different 
tracers like Open Clusters, Cepheids, HII regions, etc., see e.g. the series of papers by 
Andrievsky and collaborators from 2002 to 2005, \cite{Rudolph2006},  \cite{Pedicelli2009}, 
and \cite{Magrini2009},  among others. In the past, the metallicity gradients were
most often fitted by  straight lines in logarithm of abundance versus Galactic 
radius plots, with  slopes of about -0.07 dex/kpc. This "first order" approximation did not pose 
severe constraints to chemical evolution models, since many important  parameters of the
 models, like gas density and  infall rate of matter are supposed to be smoothly 
 decreasing functions of the Galactic radius  (\citealt{Chiappini1997}). However, evidences have
 accumulated that a single slope approximation (or two slopes, in the last reference) is not
 satisfactory, since it hides more complex slope changes that are real and should be better understood.
 For instance, the series of papers reporting observations of Cepheids by Andrievsky and collaborators
 showed that the  slope of the abundance gradient is larger than average around 5-6 kpc from the center,
 followed by  a plateau from about 6.5 kpc to 10 kpc, and then  by  a new decrease at about 10 kpc 
 (here we quote the Galactic radii as given in the original papers). As another example, 
 \cite{Twarog1997}(hereafter TAAT) analyzing a sample of  Open Clusters, have found that the Galactic
 metallicity gradient can be described  in terms of two zones with flat slopes: an inner disk with radii
ranging from 6 to 10 kpc and an outer disk for radii larger than 10 kpc. These two zones are separated by a 
 step-like discontinuity  of  about 0.3 dex in [Fe/H]. Finally, \cite{Carney2005} and 
  \cite{Carraro2007}  found that the metallicity gradient is totally flat in the outer 
 disk. All these results are in conflict with the old view of an unique and constant slope, however,
 they are not inconsistent with each other, as it will be shown in this work.

These changes of slope and step-like behavior of the gradients merit great attention since 
they provide a much deeper understanding of the disk evolution and impose real constraints to 
the existing models.  In particular, we believe that the effect of the spiral arms cannot be ignored in 
chemical evolution models, as it has most often been done. The spiral arms are the star-forming 
machines of the disk, and are the places where the youngest stars (O, B stars and  HII regions)
are found.  The massive stars are those which have the most important role in chemical evolution. 
Furthermore, when one integrates the orbits of Open Clusters in the Galactic potential for a 
time equal to their age, one finds that they were born in the spiral arms (\citealt{Dias2005}). This
is a confirmation that almost all stars are born in the arms. The spiral arms obey a number 
of rules dictated by the resonances between the epicycle frequency $\kappa$ (the frequency of oscillation
of a star around its non-perturbed circular orbit) and the frequency of rotation $\omega$ around the 
Galactic center, as observed in the frame of reference of the spiral pattern $\omega_p$. 
Among the major resonances are the inner and the outer Lindblad resonances
(ILR and OLR, or 2:1 resonances), situated where $\omega-\omega_p = \pm \kappa/2$.
According to the classical theory of galactic spiral waves proposed by \cite{Lin1964} and \cite{Lin1969},
the spiral arms are restricted to the interval between these two resonances, so that an important change
in the star-formation rate could be expected at the corresponding radii. For other theories in which 
the spiral arms are explained by the crowding of stellar orbits, the ILR and OLR have the same nature
and are equally important \citep{Kalnajs1973}. The 4:1 resonances, which appear where
$\omega-\omega_p = \pm \kappa/4$, are the equivalent of the ILR and OLR for 4 arms; in principle
a 4-arms structure could only exist between these two resonances. A stellar  orbit at a 4:1 resonance
presents 4 maxima of the radius during one revolution around the center, which gives it the
aspect of a square orbit with rounded corners.

Another major resonance is corotation, situated  where the rotation speed of the material of
the disk coincides with that of the spiral pattern ($\omega = \omega_p$). This maximizes the effect of 
the potential perturbation of the spiral arms on the stars and gas clouds in this region. The spiral 
arms are able to produce systematic radial transfer of interstellar gas with opposite directions on the two
sides of corotation, and to create a ring with a void of gas at this resonance, as it will be
discussed later in this paper.
 
A matter of concern is that it has been pointed out by several authors who analyzed the metallicity 
gradients in the Galactic disk that deviations of individual measurements from the average slope
are often found to be  larger than individual errors on the measurements (e.g. \citealt{Yong2005}, 
\citealt{Carraro2007}). This challenges the usually accepted hypothesis that the metallicity must have
a single value at any given radius (or, equivalently, that the azimuthal diffusion is very fast). 
It is important to analyze the non axis-symmetric variations of metallicity, and if they are confirmed,
 to verify if they are related to the spiral structure, or alternatively, if random local variations 
 in metallicity do exist. 

In the present paper we analyze the connection of variations in the abundances of elements with
 resonances of the spiral structure like the 4:1 resonance, corotation and OLR.
 The radii of these resonances, at least for the main spiral structure, is no more uncertain as
 they were in the past; they are now known with accuracy of the order of 0.2 kpc, as discussed by
  \cite{L'epine2010}, hereafter L+5. We adopt, for corotation, $R_c$ = 8.4 kpc, combined with a
  solar orbit with $R_0$ = 7.5 kpc. The reasons for the choice the "short scale" for
   $R_0$ are discussed in Section 3. The value of $R_c$ results from the nice agreement between 
 parameters directly determined, such as the rotation speed of the arm pattern, measured by integration
 of the orbits of Open Clusters, the radius of the ring-shaped gap in the Galactic disk (associated 
 with corotation), and the observation of the  position of the 4:1 resonance. One can find in the
 literature a number of determinations of the corotation radius that are in conflict with these measurements.
 However, we must be conscious that numerical N-body simulations, hydrodynamical simulations and
 chemical evolution models depend on a series of assumptions of unknown parameters and on simplified
 hypotheses, so that their results cannot be regarded as being as precise as those of direct measurements.
 For instance, in a recent work by \cite{Acharova2010}, a chemical evolution model gave a best fit to
 the metallicity plateau of Cepheids with corotation at 7 kpc (which would be about 6.6 kpc in the
 shorter scale adopted here). However, it was left clear in the mentioned  paper that this result 
 could not be regarded as a determination of the corotation radius. 

We are aware that dynamical measurement of the pattern speed of the spiral structure, based 
 on the location of the Hyades and Sirius moving groups in velocity space (\citealt{Famaey2005},
  \citeyear{Famaey2007}), placed the corotation at a radius $R=12$ kpc (\citealt{Quillen2005a},
  \citealt{Pompeia2011}).  This could perhaps be reconciled with the present findings in the 
  case of the existence of multiple
 spiral patterns with different pattern speeds; while the main grand-design spiral pattern
 has its corotation at 8.4 kpc, and is responsible for the plateau in the metallicity gradient
 as well as for the ring-shaped gap in the density of gas, it could coexist with an outer m=2 
 pattern whose 4:1 inner resonance smoothly connects with the corotation of the main pattern.
 This is  what happens in the  N-body simulations of \cite{Quillen2010}.
 While the corotation of the main spiral pattern would not have a large effect on the local velocity 
 distribution of stars in the solar neighbourhood, the 4:1 inner resonance of that slower outer
 pattern would be responsible for the observed substructures. The corotation of this outer spiral
 pattern would then be located at about $R=12$ kpc, near the OLR of the main structure, and could
 also contribute to a gap which seems to exist in the distribution of clusters at this radius,
 and will be discussed in this paper. We restate, however, that although we wish to keep
 open the possibility of a second pattern speed, all the analysis in this paper is based on
 what we call the main structure, with corotation at 8.4 kpc.

The accuracy now reached by direct measurements of the radii of resonances of the main  
structure allows us to perform detailed comparisons between the features of abundance 
variations and the Galactic spiral structure. This is the main focus of the present paper, which
is organized as follows. In Section 2 we analyze the expected ranges of radii covered by  orbits
of stars, to estimate how far from their birthplace young tracers can be found. This is a theoretical
background needed for a better understanding of the abundance plateaux that are discussed later
in the paper. In the subsequent two sections we present the data of the abundance tracers that  have
the most precise distances: Open Clusters (Section 3) and Cepheids (Section 4). In Section 5 
we examine a departure from axis-symmetry in chemical abundance that we call azimuthal gradient,
and we comment an extra-galactic example of non-axissymmetric gradients which looks like 
the "overlapping" gradients that we believe to exist in the Galaxy.
In Section 7,  from the observation of the step-like decrease of metallicity at corotation,
we derive a lower limit to the life-time of the spiral structure.

\section{Range of radii swept by stars}

We review here the basic rules that regulates the range of radii swept by
the orbits of stars or Open Clusters in the Galaxy. 

A typical orbit is shown in Figure 1. This
orbit was obtained by numerical integration for a star with initial velocity 40 km/s in excess 
with respect to circular rotation at the initial radius. If we are interested on the radial
motion only, there is a classical solution which consists in introducing an effective potential
governing the radial motion of a particle of
a given angular momentum $\ell$ in a central force field. The effective potential is the
sum of the original central potential plus a  rotational energy given by $U= \ell^2 /2mr^2$,
where $m$ is the mass of the particle and $r$ its distance from the center. Figure 2 shows
two examples of effective potential curves, with different angular momentum of the star. The Galactic potential
is obtained by integration of the rotation curve of the Galaxy (the gravitational force is equal
to $mV(r)^2/r$, where $V(r)$ is the rotation curve). We arbitrarily set the potential equal to zero
at a large distance (we used 60 kpc). Note that adding an arbitrary constant to
the potential energy curves does not change the dynamics of the stars, since we are not investigating
the escape probability and we restrict our study to orbits with small perturbation energy. In the figure,
a dot represents the radius of minimum energy (circular orbit), and the
dashed lines show the range of radius swept by the star for a not too large perturbation. 

\cite{L'epine2008} presented an histogram of the perturbation velocities of the Open Clusters
(the residual velocity after subtracting the velocity of the rotation curve). The histogram
presents a peak at about 15 km/s, and extends to about 50 km/s. The ranges of radii illustrated
in Figure 2 were computed with a perturbation velocity of 40 km/s, so that they are close to the 
maximum observed range. The most important result is that the stars travel between their normal maximum
and minimum Galactic radius in a time shorter than one revolution around the Galactic center.
Another mechanism capable of producing changes in the Galactic radius of stars is the scattering
by the corotation resonance. This was investigated in detail by \cite{Sellwood2002} 
and \cite{L'epine2003}, showing that the scattering is able to impart larger radial amplitudes,
but it  requires a longer time to be effective, compared to the normal epicycle oscillations 
described previously.

\begin{figure}
\includegraphics[width=64mm]{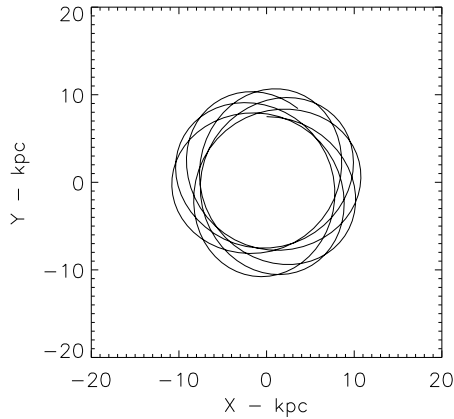} \caption{Typical orbit of a star in the Galactic potential,
with minimum radius of about 7 kpc and maximum of about 11 kpc.}
 \label{fig1}
\end{figure}

\begin{figure}
\includegraphics[width=64mm]{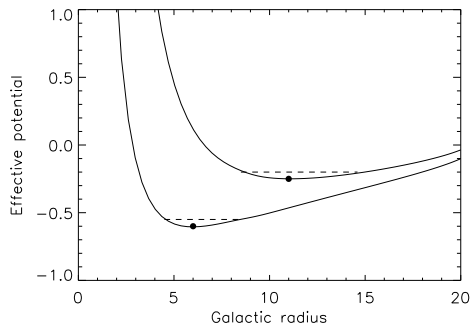} \caption{The gravitational potential of the Galaxy
with a centrigal barrier. Two curves are shown, corresponding to minimum energies (circular
rotation) at 6 and  11 kpc.
The radial position of circular rotation is indicated by a dot. The radial range of the stellar orbits
for a small energy perturbation is indicated by dashed lines.}
 \label{fig2}
\end{figure}

\section{Gradients based on Open Clusters}

Our analysis of  metallicities of Open Clusters is based on two different samples, 
which were constructed using different criteria to select the results from the literature.
The first sample only includes [Fe/H] measurements obtained 
with high-resolution spectroscopy. This is basically the  same sample of 45 object collected
by \cite{Magrini2009}, updated with a few recent observations. These include
NGC 6633, NGC 2539, NGC 2447, IC 2714 and NGC 5822 \citep{Santos2009}, NGC 7160 \cite{Monroe2010},
CR110, NGC 2099, NGC 2024 and NGC 7789 \citep{Pancino2010}, Tombaugh2 \citep{Villanova2010}.
Two measurements presented by \cite{Gratton2000}, of IC4725 and NGC6087, although not so
recent, were added to this sample.

The second sample contains metallicity derived from spectroscopy as well as from photometry and
other methods. These data are less reliable, and their use is  justified by the larger
number of measurements available (131 compared to 58 for the first sample). This second sample
 is based on the on-line version of the {\it New Catalogue of Optically visible Open Clusters and
 Candidates} published by \cite{Dias2002}, and often updated\footnote{Available at the web page
 $www.astro.iag.usp.br/{\sim}wilton$}. This catalog contains a list of 153 clusters with available
 metallicity measurement, and gives the adopted reference for each  entry, as well as the
 method of observation and estimated error. The metallicity indicator used is [Fe/H], usually given in the
original references. However, in a few cases the metallicity included in the catalog  is $Z$, 
from which the Fe abundance can be recovered using the relation [Fe/H] = log (Z/0.019)
 (see \citealp{Carraro1999}).

 Minor changes or updates were performed for a number of objects, often giving preference 
 to metallicity and distance from the same author. For NGC 2168 we adopted the
 distance given by \cite{Pohnl2010}(0.8 kpc); for NGC 2489, the distance given by
 \cite{Piatti2009} (1.8 kpc); for NGC2335, the same distance used by TAAT (1.15 kpc);
 for NGC188, the distance given by  \cite{Meibom2009} (1.77 kpc), and for NGC2360, the distance given by 
 \cite{Hamdani2000} (1.09 kpc); for Berkeley 20 and Collinder 261, we used the data of \cite{Sestito2008};
 for Berkeley 31 the distance by \cite{Yong2005} (5.3 kpc); for IC4756, NGC 2682 and 
 NGC 3680, the [Fe/H] data of \cite{Santos2009}. 
 
 In the analysis of this second sample, we only make use of data with errors smaller 
 than 0.2 dex in [Fe/H]. A number of authors do not give an error estimate (see the on-line 
 catalog), but we adopt an error estimate of 0.1 dex based  on the similarity of the technique 
 with other work. Data of \cite{Ann2002} are an exception, for which we believe that the error 
 is larger than the limit adopted here, because some discrepancies were found in two Open Clusters
that are in common with \cite{Hasegawa2004}, and because for some clusters it seems
that $Z$ is given instead of [Fe/H], which is not clear. The data from  Hasegawa et al.,
(not included in the on-line catalog), on their turn, are not included 
because the error estimated by the authors is 0.3 dex. The error on the metallicity
of Berkeley 21 (the Open Cluster with the lowest metallicity) is about 0.3 dex \citep{Tosi1998}, 
and that of NGC6451 is about 0.5 dex \citep{Paunzen2003}. It should be noted that
our procedure of eliminating objects with large errors on metallicity has affected more
severely objects with  [Fe/H] $<$ -0.5; it may happen that our sample is biased 
in showing only few  clusters with very low metallicity. Note that there is a large
 overlap in the list of clusters between the high-quality sample (first sample) and this one.

For both samples (high-resolution, and general) we re-computed the galactocentric distances
of all the objects using $R_0$ = 7.5 kpc  instead of the IAU recommended value 8.5 kpc. 
This is a more realistic estimate of $R_0$, as for instance the (preferred) Models 1 and
2 from Table 2 of \cite{McMillan2010} assume $R_0 = 7.3$ and 7.8 kpc respectively. A higher
value of $R_0$ would imply an unrealistically high circular speed. See also the recent work 
of \cite{Sofue2011}. But the use of this shorter scale is not something new in the literature,
since many specialists in the Galactic distance scale recognized that $R_0$ is about 7.5 kpc; see  the 
reviews by \cite{Fich1991} and by \cite{Feast2008}, or the papers
 by \cite{Glushkova1999} and others  of the same group. It is preferable to use this
value of $R_0$ when precise  distances in different directions are to be compared,
like for instance when stellar distances are compared with the position of resonances.

In Figure 3 we present [Fe/H] as a function of galactocentric distance, for the  sample of 
Open Clusters with high-quality measurements.  The data of the second sample (lower
quality but larger sample) are presented in a similar way in Figure 4. In both figures
different colors were used to indicate young, intermediate age and old Clusters. In
Figure 5 the same data of Figure 4 (larger sample)is presented
with histograms on the two axes. 
 
\begin{figure}
\includegraphics[width=84mm]{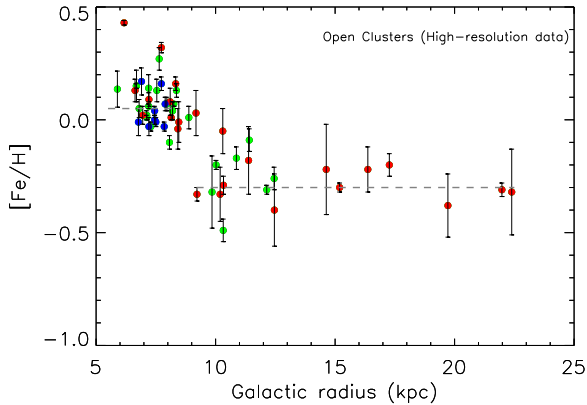} \caption{Fe abundance of Open Clusters normalized to solar as
a function of radius measured by different authors. Only results of high-resolution spectroscopy
are presented (our Sample 1). The errors on [Fe/H] are taken from the original references. 
Galatocentric distances were re-calculated for $R_0$= 7.5 kpc. Different colors correspond to ranges
of ages of the Open Clusters; blue: age $<$ 200 Myr, green: 200$<$ age $<$ 1200 Myr, red: age $>$ 1200 Myr.
The dashed horizontal lines indicate the average metallicity on both sides of corotation.}
 \label{fig3}
\end{figure}

\begin{figure}
\includegraphics[width=94mm]{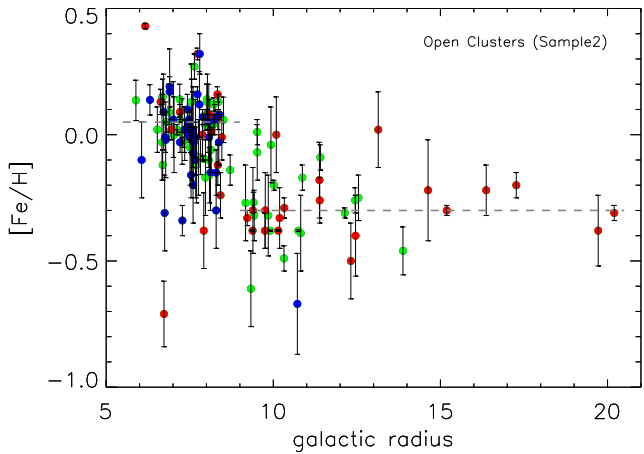} \caption{Like Figure 3 for our  Sample 2
of Open Clusters, which includes [Fe/H] results of different quality, based on spectroscopy,
photometry and other methods. Measurements with errors larger than 0.2 dex were excluded.}
 \label{fig4}
\end{figure}

\begin{figure}
\includegraphics[width=84mm]{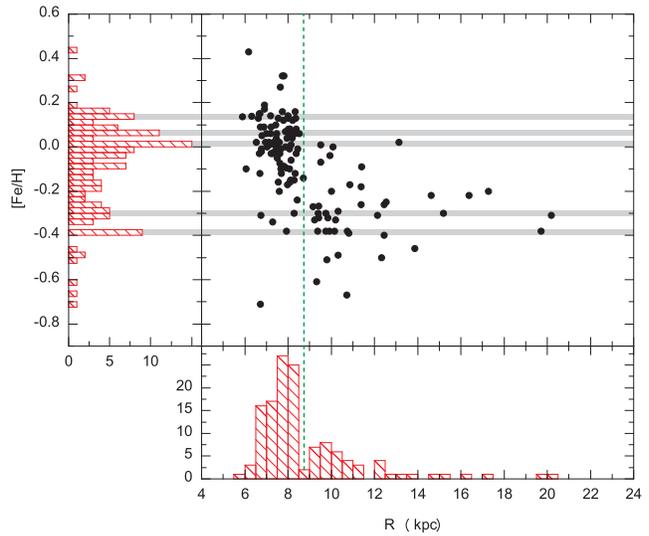} \caption{The same data of Figure 4, without
error bars and without age information.
On the two sides of the plot, the histograms of the data on the corresponding axis are presented.
The green line indicates the corotation radius, which is a frontier between two distinct regions of the disk
and also corresponds to a minimum of star formation rate.  The grey bands  emphasize
 concentrations of clusters at metallicities corresponding to  peaks in the histogram on the left side.} 

 \label{fig5}
\end{figure}

\subsection{The step and the gap at corotation}
The first remark to be done is that the step in the Fe abundance first observed by TAAT is confirmed;
it appears clearly in Figure 3 which presents  high-quality data that were not available at the
epoch of TAAT's paper. The height of the step is about 0.3 dex. The Galactic radius at which it is
observed (8.5 kpc) is the same found by TAAT after correction for the different values of $R_0$.
In addition to the step, one can  see a gap in the distribution of clusters at Galactic radius 8.5 kpc, which is
clearer in Figure 4 and in the lower histogram of Figure 5, due to the larger number of Clusters plotted
in these figures. The radius of the gap coincides with corotation and with the Cassini-like ring-shaped 
region void of gas discovered by \cite{Amores2009}. The gap in the distribution of clusters tells us that
the star formation is hindered in the region of gas void, while the step in the metallicity distribution
indicates that  the gas void  forms a barrier that has prevented the transportation of gas from one
side to the other, so that the metallicity evolved independently  on the two sides.

\subsection{The flatness of the external gradient and cluster ages}

The metallicity distribution as a function of radius confirms that the  metallicity gradient is flat in 
the outer part of the disc,  a fact which was observed by many authors, among them TAAT,
 \cite{Carney2005}, \cite{Yong2005}, \cite{Carraro2007} and \cite{Magrini2009}. There was not yet
a consensus on where the flat part begins, but from Figures 3 to 5  it seems clear that the radius at which
the "base level" of  metallicity of the outer disc starts is the corotation radius.
 
In Figure 3 and 4 we separated the Open Clusters in 3 groups of age, the young ones (age $<$ 200 Myr), intermediate
(200 $<$ age $<$ 1200 Myr) and old (age $<$ 1200 Myr). One can see that there is no young cluster with 
[Fe/H] smaller than about -0.3 dex, on the inner side of corotation. The young clusters present  are
 IC 2581, NGC 2343 e NGC 6716. Surprisingly, many metal-rich clusters with [Fe/H]  about 0.1 dex are observed 
 in the same range of Galactic radius. This indicates that large metallicity differences are present in the 
 gas  situated at different azimuthal directions in a same interval of Galactic radius, which suggests that
 a fast mixing of the gas in the azimuthal direction, usually assumed in chemical evolution
 models, is a poor approximation.
 
 There is only one young cluster  beyond corotation in Figure 4, Dolidze 25.
 This brings the question: at what Galactic radius does star formation stop? Star formation tracers 
 like HII regions, compact HII regions revealed by  CS emission and other tracers are seen up to
 about 14 kpc, e.g. see L+5 and \cite{V'azquez2008}. In  Figure 3 and even in Figure 4 we are working with
 small samples of objects with well measured metallicity, so that we should not conclude from these
 samples that young clusters do not exist at larger radii. 
 
 In Figure 6 we examine the age versus Galactic radius relation of the larger sample of clusters
 (1266 objects extracted from the web page of \cite{Dias2002}) which have measured distance and age,
 without any other requirement (for this study the existence of high-resolution spectrometry
 is not so relevant). This figure confirms that young clusters, if they exist, are very rare beyond
 14 kpc. The most distant object among the young clusters (in the lower part of 
 the Figure 6) is Teutsch 45, at Galactic radius 14.3 kpc, which presents, however, a large error
 on distance ($\pm$ 1.6 kpc). Other objects which were previously believed to have similar Galactic radii,
 like Dolidze 25 and NGC 1893, had their distance  corrected based on recent publications (\citealp{Delgado2010}, 
 \citealp{Prisinzano2011}). We note however that Carraro et al. (2010) report the detection 
of diffuse groups of young stars in the third quadrant of the Galaxy, at distances between 14 
and 19 kpc (converted to our short scale). The estimated distance of VdB-Hagen 4, which contains 
B stars, is of the order of 19 kpc. Similarly \cite{Moitinho2006} detected "blue plumes", which are 
groups of young stars up to distances of 16 kpc (in our scale) in the third quadrant of the Galaxy
(the other "normal" clusters listed by the same authors are already included in the cluster database that
we presented).

The flatness of the gradient of the outer disk (Figures 3 and 4) deserves an explanation. One possibility is the flow of gas 
from corotation to the external regions, which is a consequence of the interaction of the gas  with the
spiral potential perturbation. The radial flow is  predicted theoretically and confirmed by 
hydrodynamical simulations (\citealp{Mishurov2000}, \citealp{L'epine2001}). This flow  transports the relatively 
 high metallicity gas of the corotation zone towards outer regions, and flattens the gradients. 

Another possible reason for the flatness of the gradients is that a large fraction of
the clusters that are presently observed at galactocentric distances of 15 to 20 kpc were born 
at smaller distances. This is naturally expected based on the discussion presented in Section 2. As we
move towards larger radii, the curves representing the effective potential (Figure 2) become 
shallower, and for the same perturbation energy, the clusters cover a larger Galactic radius range.
This hypothesis aims to explain why there are no (or almost no) young clusters beyond about 14 kpc.
The conditions for star formation (probably, the gas density) are not fulfilled at large radii; 
the age of the clusters situated there is sufficiently large to allow them to be born in an inner radius
and to have traveled to their present position. An additional argument in favour of the hypothesis 
that the outer clusters are travelers born in inner regions is their distribution in Galactic 
component $z$ (direction perpendicular to the Galactic plane), shown in Figure 7.  Simple 
integration of orbits of stars (or clusters) in the Galactic gravitational potential shows that if
they are launched from a radius of 12 kpc, when they reach a radius about 18 kpc, their  $z$ coordinate
can easily  reach values of the order of 2 kpc, similar to the observed scale height. 
In such experiments the same initial velocity perturbations in the $z$ direction required to explain
the local (solar neighbourhood) scale height is used. Since the gravity of the disk decreases with Galactic 
radius, the amplitude of the oscillation in the $z$ direction increases, if a star moves outwards. 
The good match of the scale-heights points towards a common origin of the velocity perturbations.
The smooth increase of average age with distance seen in Figure 6 is also an argument in favour of 
the "travelling clusters" hypothesis. Note, however, that there are  indications that the two outermost
old Open Clusters, Berkeley 29 and Saurer 1, are of extragalactic origin \citep{Carraro2009}.

 \begin{figure}
\includegraphics[width=84mm]{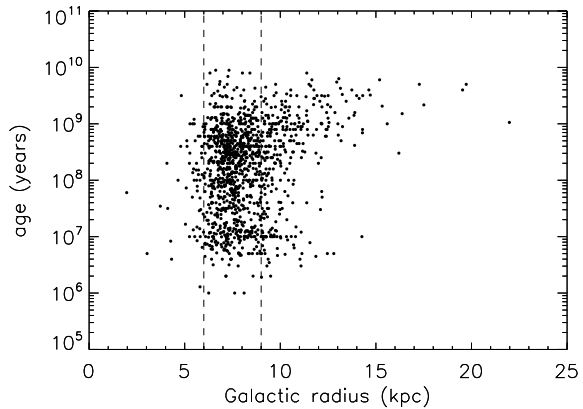} \caption{The age of Galactic clusters as a function
of Galactic radius, for the larger sample of clusters which have estimated distance and age
(with no need for metallicity measurements). The two vertical dashed lines are at symmetrical
Galactic radii with respect to the Sun (at 6 kpc and 9 kpc) to emphasize that the cluster distribution 
is not symmetrically distributed around the Sun.  }
 \label{fig6}
\end{figure}  

\begin{figure}
\includegraphics[width=84mm]{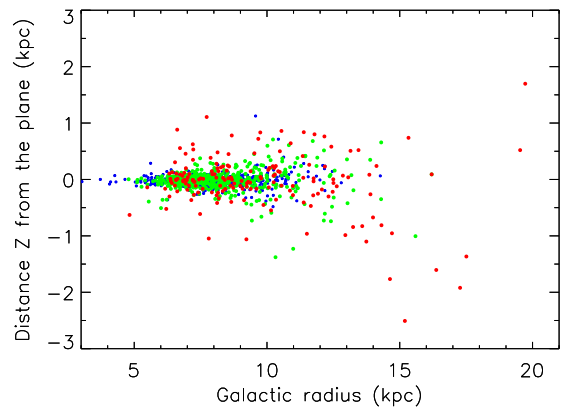} \caption{The distribution of Galactic clusters as a function
of Galactic radius and of distance Z from the Galactic plane, for the larger sample of clusters which have
estimated distance and age. The ages of the clusters are indicated by colors, like in Figures 3 and 4.
The blue color symbols (young Clusters) are densely concentrated near the Galactic plane, but are hidden 
by the green symbols. Note that the distance scales are different on the two axes. The older clusters reach 
larger $|Z|$ and larger $R$ values.}
 \label{fig7}
\end{figure}

\subsection{A possible zone of enhanced star-formation efficiency between the 4:1 resonances}

Does the 4:1 resonance have any visible effect on the star-formation rate? This resonance, which 
is the ILR of the 4-arms structure, as previously discussed, is  at 6.2 kpc, see L+5.
Since our Galaxy has a strong 4-arms component, the frontier of the region where 4 arms are allowed to exist
with the region with only 2 arms  is in principle an important boundary. The star-formation machine (the spiral arms)
and the mechanism that produces the inflow of gas have their power increased beyond  that radius
due to the larger number of arms. All the higher resonances, some of them possibly having associated
spiral arms, are situated between the 4:1 resonance and corotation. Indeed, both the lower histogram
in Figure 5, and Figure 6, which exhibits a larger sample of clusters, show a strong increase 
in the density of  clusters at about 6.2 kpc.
Of course, the interstellar extinction could be a cause of the cut-off in the counts of clusters
at Galactic radius about 6 kpc. However, the 4:1 resonance is only 1.3 kpc from the Sun, 
which corresponds to a visible extinction a little larger than 1 magnitude in the direction of 
the Galactic center. In the opposite direction, many clusters are observed at Galactic radius 
12 kpc, which is much farther from the Sun (5.5 kpc). Therefore, the abrupt increase in 
the density of clusters beyond 6 kpc  seems to be real, not an effect of extinction.
It can be seen in Figure 6 that there are much  more Clusters on the right side of the
vertical line at 9 kpc than on the left side of the line at 6 kpc, although the two lines are 
equidistant from the Sun.  
The outer 4:1 resonance ($\Omega_p = \Omega + \kappa/4$) is situated at 12 kpc. For the same reasons 
we could expect a decrease of the star formation rate and cluster counts beyond that radius. 
However, the frontier does not seem to be so well defined in this case. The probable reason is that the 
completeness of the sample of clusters is smaller, at distances of about 5.5 kpc.
Nevertheless, it seems that the Galactic radius interval 6 to 12 kpc, the region between the 
two 4:1 resonances, except for the corotation gap, is a privileged zone for star formation.
As we discussed, the old clusters situated at large distances possibly originated in the 6-12 kpc zone,
and also most of the intermediate-age clusters situated in the neighbouring region 12-14 kpc could possibly 
be born inside the 12 kpc radius.
 
 At 12 kpc  there is some indication of a minor gap in the distribution of clusters (see eg 
 the histogram in Figure 5). In the  alternative interpretation of the spiral structure discussed in the 
 introduction, in which a secondary pattern speed coexists with the main one, 12 kpc is the radius 
 of corotation of the outer pattern. Some of the properties of the distribution of clusters could 
 also be explained by this interpretation.
 
 \subsection{Fine structure of the metallicity distribution} 

Another interesting aspect of the metallicity gradients shown in Figures 3 to 5 is that they
present concentrations of clusters around a number of horizontal (constant metallicity) lines.
Some of the peaks of the histogram on the left side of Figure 5 seem to be significant. For
instance the peak at 0.15 dex also corresponds to a group of Clusters with the same metallicity
in the sample with only high-quality measurements shown in Figure 3. 
We prefer to postpone the discussion on concentrations of metallicities around preferential values to the
later section on Cepheids, since the metallicities of these stars, which form a completely independent 
sample, confirm our statement.  We remark that in Figures 4 and 5 several
old clusters situated inside the corotation radius  have high metallicities ([Fe/H] $\approx$ 0.15 dex),
which are larger than the average value for young clusters. This poses the interesting question: is 
the metallicity in the solar neighbourhood decreasing with time (more recent clusters have
smaller metallicities) or did the old clusters situated around the solar radius have born in
inner regions of the Galaxy? Still another possibility would be to associate this observation with
the azimuthal differences in chemical abundances, that are disussed in a later section.
This question shows again the importance of taking into account the radial motion of the tracers to understand the 
metallicity gradient. 

In Figure 6 one can see a concentration of clusters with age $10^7$ years; this can be interpreted  as a rounding
effect in the papers reporting ages, rather than an evidence for a star-formation episode. It should be remembered
that the large sample presented in that figure (1266 objects) contains data with non-uniform quality. 
 
We will not discuss here the abundances of $\alpha$-elements in Open Clusters, as only a small number of clusters
have been measured. The $\alpha$-elements can be better analyzed based on the sample of Cepheids, discussed
in the next section.

\section{The gradients based on  Cepheids}

An impressive set of metallicity data for 130 Cepheids, with abundances derived for many elements
based on  high resolution spectra, is presented in the series of papers by  Andrievsky and 
collaborators: \cite{Andrievsky2002a, Andrievsky2002b, Andrievsky2002c},
 \cite{Luck2003}, \cite{Andrievsky2004}, \cite{Kovtyukh2005}, \cite{Luck2006}.
 These data constitute the main source for the present analysis. The range of Galactocentric distances
covered by the sample is  4 to 16 kpc.

We revised the distances of the Cepheids to the Sun given in those papers as next explained,
and subsequently re-computed the galactocentric distances using $R_0$ =7.5 kpc, as it was done 
for the Open Clusters. The stellar distances in the original papers on metallicity measurements were
usually taken from the catalog of \citet{Berdnikov2000} with on-line updates. We
re-estimated them using the expression for intrinsic colors as a function of period given
by \citet{Abrahamyan2003} and in addition, using  R$_v$=A$_V$/E(B-V)= 3.5 instead of 
the usual  3.1 used by Berdnikov et al.. The reason for this change is that R$_v$ depends 
on the spectral type of the stars (a correction must be made to the effective wavelength of the 
B and V filters when they are convolved with the energy distribution of the stars (see 
\citealp{Avzusienis1969}). L+5 found that the distances of the Cepheids 
calculated this way were better  correlated with the kinematic distances than the original 
distances given by Berdnikov et al.

These minor corrections of distances do not change the main picture of the gradients
that emerged form the series of papers of Andrievsky and collaborators, that we described
in the Introduction. 

\begin{figure}
\includegraphics[width=84mm]{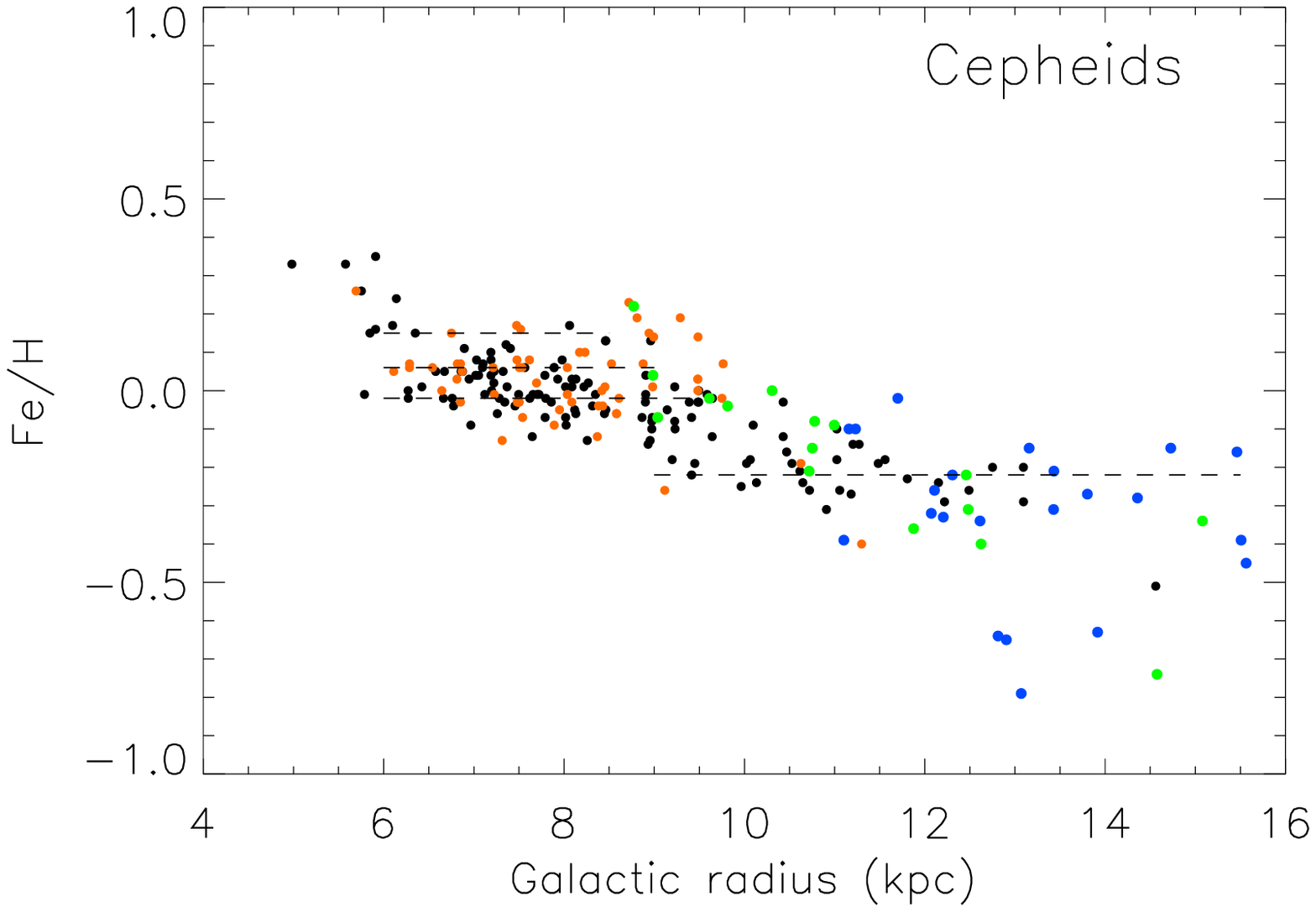} \caption{Fe abundance of Cepheids normalized to the 
solar values as a function of Galactic radius measured by different authors: black dots, the series of papers by
Andrievsky and collaborators, orange dots, {\citet{Luck2006}}, blue dots,{\citet{Yong2005}}, and
 green dots, {\citet{Kovtyukh2005}}. Minor corrections have been applied to the data of some of
these authors (see text). The horizontal dashed lines emphasize  metallicities around which
the Cepheids seem to be concentrated. The galatocentric distances are based on $R_0$ = 7.5 kpc.}
\label{fig8}
\end{figure}

We present in Figure 8 the Fe abundance in Cepheids as a function of Galactic radius.
 In this plot we made small corrections to the values of [Fe/H] given in two papers.
 The papers of Andrievsky et al. and Luck et al. are based on the use of the same approach
 of spectroscopic analysis (the same system of oscillator strengths and the same methods of 
determination of the stellar fundamental parameters) so that the results are directly comparable.
In spite of this, a minor mismatch appeared in the average metallicity for a large number
of stars in the same galactic radius (there were no stars in common). We added 0.03 dex to
 the measurements of \cite{Luck2006} to correct for the difference in the average value of [Fe/H],
 considering the data of Andrievsky et al. as the correct ones.
 Another correction, of larger magnitude, was applied to the data of \cite{Yong2005}.
 Their sample contains two stars in common with the series of Andreievsky et al: NT Pup and CE Pup,
 for which Yong et al. obtained [Fe/H] values lower by 0.25 and 0.2 dex respectively.
 The average difference for the stars situated in the region of overlap of the two samples
 was 0.3 dex. We added 0.25 dex to the [Fe/H] results of Yong et al. to match the data
 of Andriesky et al. This correction has an uncertainty of the order of 0.05 dex.
 The sample of \cite{Kovtyukh2005} contains 4 stars in common with Andrievsky et al.:
 RX Aur, RW Cam, T Mon, SV Mon; the average metallicity difference for these stars
 is negligible.
 
At a first glance, the traditional inclined straight line used in the past to fit metallicity
gradients as discussed previously, would seem to be justified, to describe the ensemble of 
Cepheid data. However, a step function 
(or Heavyside function), not shown, with a step of 0.3 dex at 9.5 kpc, produces a fit of
the same quality (Chi-square divided by the number of degrees of freedom), if we restrict the fit to the 
interval 6 to 13 kpc. In other words, this plot cannot be used  to dismiss the existence of a 
TAAT-like step in the sample of Cepheids. The magnitude of the step, of about 0.3 dex, 
seem to be a little smaller than the one found for Open Clusters, but the difference
is within the errors.

The gap in the distribution of Galactic radii of the Cepheids, similar to that of the 
Clusters, can be seen at about 8.7 kpc. It seems to be narrower and a little displaced with
respected  to the corotation radius (8.4 kpc). One explanation could be that the Cepheids
are on the average older than the young  Clusters and had more time to partially invade
the corotation gap. The range of ages of the Cepheids is about 50 to 300 Myr
 \citep[e.g.][]{Bono2005}. Another possibility is that the errors in distance are larger
 for Cepheids.

A number of horizontal lines are plotted in Figure 8 at [Fe/H] values  which seem 
to concentrate a large quantity of stars. Some of these lines  coincide with the 
[Fe/H] values with relatively larger density of  clusters in Figures 3 to 5, like
for instance at 0.0 dex and 0.15 dex. This is not surprising, since the two samples are
constituted of  populations of  young objects; both can be considered to represent
the chemical abundance of the interstellar gas at their birthplace at the present epoch,
since 200 Myr (the order of magnitude of the ages) is a small interval of time for the chemical
evolution of the disk. 

To produce the horizontal alignments of Cepheids seen in Figure 8,  there must be relatively
few regions in the Galactic plane with  active star formation. In any one of these  regions,
stars are formed from the local gas with a same metallicity, and then start their travel 
in radial distance from the Galactic center according to their initial perturbation velocity. 
The gas of the disk possibly presents a smooth variation of metallicity with radius, except 
for the step at corotation. If there were star forming regions distributed all over the Galactic plane,
with a smooth distribution of metallicity, it would not be possible to distinguish preferential
metallicities in the distribution of clusters. Certainly the Cepheids are born in spiral arms,
like the  clusters (in the case of the clusters this was shown by direct integration of their
orbits for a time equal to their age, see \cite{Dias2005}. Therefore, the spiral arms
are the obvious candidates to be the discrete star-forming regions required to explain the
existence of peaks in the metallicity distribution.  Then, a natural question is: many arms
are elongated structures, spanning a range of Galactic radius, why don't they  produce 
a broad distribution of metallicity that would smooth out the horizontal lines seen in Figure 8?
One possibility is that some portions of the arms, like the
corner of the square-shaped resonant arm discussed by L+5, are more efficient 
star formation regions than the others, and dominate. Another possibility is that due to the flow
of gas along the arms the gradients of metallicity are small within a given arm.

 \subsection{The $\alpha$-elements}

In order to reduce the errors on the abundance of $\alpha$-elements,
 we averaged five elements. These are  O + Si + S + Mg + Ca, taken
 with equal weights. We shall designate them as $\alpha$-elements
for simplicity. The choice of these elements was based on the fact they are classical
$\alpha$-elements, and they were measured by different authors who performed spectroscopy of 
Cepheids.  Since O was not measured by \cite{Yong2005}, for the data of these 
authors we calculated the average of the remaining elements. The results are shown in Figure 8.
We also performed minor corrections to the data presented in this figure, to match of the alpha
abundance data of different authors: the corrections were -0.04 dex for \cite{Luck2006}
and -0.1 dex for \cite{Kovtyukh2005}.
 
The step-like decrease in the abundance of $\alpha$-elements at corotation looks  
 similar to that of the Fe abundance.  We plotted two horizontal lines in the Figure,
 to indicate the average metallicity inside corotation and the "base-level" beyond corotation.
 The difference between them is of 0.25 dex, a little smaller than the step  in Fe abundance
 but the difference is not significant compared to the scattering of the points.
 
\begin{figure}
\includegraphics[width=84mm]{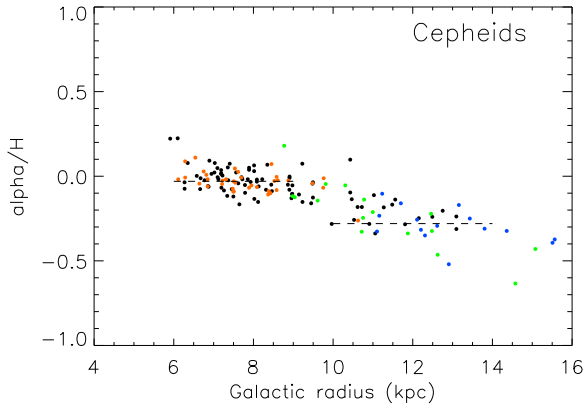} \caption{Abundance of $\alpha$-elements 
(average of O + Si + S + Mg + Ca) in Cepheids normalized to the solar value, in dex, as a
function of Galactic radius ($R_0$= 7.5 kpc) measured by different authors. The colors are the 
same as in Figure 7. The dashed lines indicate the average abundance of 
$\alpha$-elements in regions inside corotation (-0.03 dex) and the baselevel at large distances
(-0.28 dex).}
\label{fig9}
\end{figure}

\subsection{The [$\alpha$/Fe] ratio}

The obtained [$\alpha$/Fe] ratio, where $\alpha$ is composed of the same elements   above,
is presented in Figure 9. The same minor corrections that were applied independently to
Fe and $\alpha$-elements abundance were used to compute [$\alpha$/Fe]. The radial
distribution shown in Figure 9 is surprisingly flat, showing an homogeneous abundance 
pattern all over the disk.

 Although our discussion here has its focus on Cepheids, we have to
 make comparisons with results for different objects. Our results seem to be in 
 contradiction with those from \cite{Carney2005},  who have observed a relatively large (0.2 dex) 
 increase of the [$\alpha$/Fe] ratio beyond 10 kpc. The origin of this discrepancy
 is mostly due to the Fe abundance, found by Carney et al. to be about -0.5 dex in the outer disk,
while \cite{Andrievsky2002c} derive about -0.2 dex, not much different from
TAAT ($\sim$ -0.3 dex).  Supposing that our results based on Andrievsky's data are correct, 
the uniformity of the [$\alpha$/Fe] ratio with Galactic radius is an argument against
 the hypothesis raised by \cite{Yong2005}, according to which the outer Open Clusters could be 
 associated with mergers, and consequently, have very different formation histories from those
 born in our Galaxy.

\cite{Acharova2010} presented in the introduction of their paper a detailed discussion of the 
problem posed by the similarity of the Fe and $\alpha$ abundances. According to the traditional 
view, most of the Iron is synthesized in type Ia supernovae (SNe), whose progenitors are stars 
with ages of several billion years. On the other hand, $\alpha$-elements are synthesized 
in type II SNe, associated with massive short-lived stars. When star formation begins
somewhere, the abundance of $\alpha$-elements grows quickly, while the Fe abundance
only starts to increase much later. However, more recently \cite{Matteucci2006} developped
a new formulation for the Type Ia SNe rate, in which up to  50 per cent of the total
Type Ia SNe should be composed by binary systems with lifetimes as short as $10^8$ yr.
Taking into account that type II SNe also produce Iron, at least 2/3 of the Iron is produced in
a fast process. Consequently, the [$\alpha$/Fe] ratio is not very sensitive to the 
time elapsed since the beginning of star formation, and is a good indicator of
a recent star formation history. In this interpretation, the flatness of the distribution
in Figure 9 is not so surprising.

Still, it is possible to obtain some insight on the star formation history from this figure.
One can see a number of stars with [$\alpha$/Fe] of the order of 0.1 dex, in the range 
of galactocentric distances about 7 to 13 kpc. This is an indication that there has been
intense star formation activity in recent times in this Galactic radius range.
Interestingly, the stars with high metallicity  seen around 6 kpc in Figure 8 do not have
a high [$\alpha$/Fe] ratio.

\begin{figure}
\includegraphics[width=84mm]{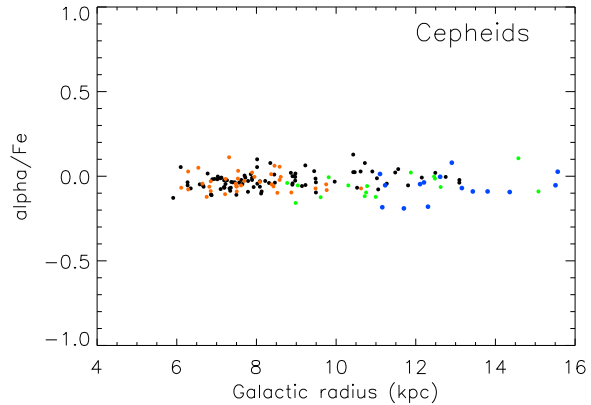} \caption{[$\alpha$/Fe] as a function
 of Galactic radius. The color scheme is the same of  Figure 7.}
\label{fig10}
\end{figure}

\section{The azimuthal gradient}

We show in Figure 11 the azimuthal distribution of Fe abundance 
in the sample of Cepheids.  This angle is the galactocentric angle
in the Galactic plane, defined to be $0^o$ in the direction of the Sun
and positive in the direction $\ell = 90^o$.  Obviously, if we could observe all the $360^o$,
no average gradient would be seen. In this example we restricted the range of Galactic radius to 
go from  7 to 11 kpc. Very similar plots can be obtained if we use the abundance of $\alpha$-elements
instead of [Fe/H], or if we change the range of radius. 
Note that if we transform the two extremities of the line segment fitted to the data
in the figure into points on the Galactic plane, their distance is about 6 kpc,
taking the average Galactic radius equal to 9 kpc. Therefore, the azimuthal
gradient transformed to distance units would be of the order of 0.05 dex/kpc,
which is not negligible compared to the radial gradients found in the literature. 
For instance, \citet{Pedicelli2009} have found an overall radial gradient of 0.05 dex/kpc.

We attribute this local azimuthal gradient to the spiral structure. For example, there
is an excess of on-going star formation  in the second Galactic quadrant  
($\ell = 90-180^0$) compared to the third one, due to the presence of the Perseus star 
forming region. This can be seen in maps of the spiral structure based on young tracers, 
like those from \cite{Hou2009} and L+5. 

The main purpose of presenting
this azimuthal gradient is to call attention to the risk of simple interpretations of 
radial metallicity gradients, and to emphasize the role of non axis-symmetric spiral arms.
Obviously the usual assumption in chemical evolution models of almost instantaneous mixing of the
enriched gas in azimuth is a poor approximation. It must be kept in mind that within
1 kpc  from corotation (which includes the solar vicinity), the differential velocity
of the gas with respect to the spiral structure is so small that it takes more than
1 Gyr for the gas to complete a turn around the Galactic center, in the frame of reference
of the spiral structure.

\begin{figure}
\includegraphics[width=84mm]{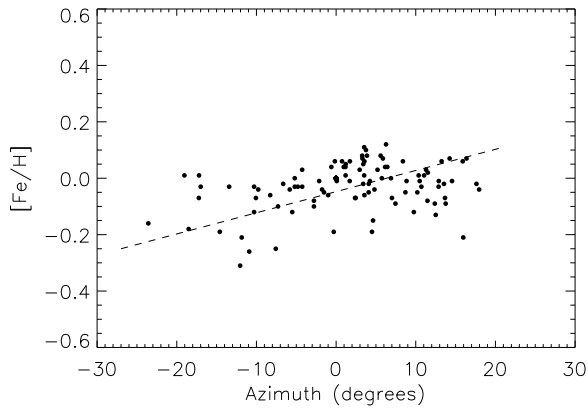} \caption{The Fe abundance  of the sample of
Cepheids as a function of the azimuthal angle. This angle is the galactocentric angle in the 
Galactic plane, defined to be $0^o$ in the direction of the Sun and positive in the 
direction $\ell = 90^o$. Only the Cepheids in the range of Galactic radius from 7 to 11 kpc
are included.}
 \label{fig11}
 \end{figure}
 
 \subsection{The metallicity gradient step  in M83}

Since there is a widespread concept that abundance mixing in the azimuthal direction
 (along a circle around the Galactic center) is very efficient, we would like to 
 present an additional example in which the simple  parameterization of abundance patterns as
 a function of Galactic radius only is a poor one, and to show that
 there can be more than one "organized" (in contrast with simple noise or random distribution)
 metallicity at a given radius. Figure 12 presents the gradient of Oxygen abundance in M83, reproduced
 from \cite{Bresolin2009}. Coincidentally, the step occurs at about the same distance than in our Galaxy.
 Where the step appears we can find in M83 "overlapping" gradients which resemble the ones we can see in Figure 3.
 We do not claim, however, that  8.5 kpc is the corotation radius in the case of M83, since there are many
 discrepant determinations of this corotation in the literature.

\begin{figure}
\includegraphics[width=84mm]{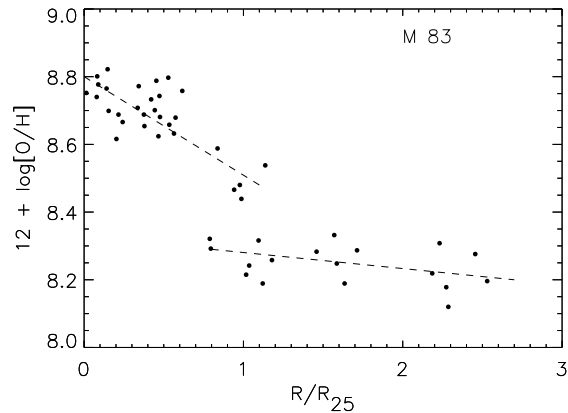} \caption{The Oxygen abundance as a function of radius
in M 83 (NGC5236) measured by (\citealt{Bresolin2009}). The galatocentric distances are in units 
of $R_{25}$, equal to 8.4 kpc according to the authors. The tracers are HII regions.}
 \label{fig12}
\end{figure}

\section{A lower limit to the age of the spiral structure}

We have shown that the step in chemical abundance first noted by TAAT and confirmed
in the present work coincides with the corotation radius; the step is explained it by the
independent chemical evolution on the two sides of the barrier 
constituted by the ring void of gas  and by the "pumping out" mechanism.
On each side of the barrier, there is exchange of gas between neighbouring Galactic radii
due to winds of massive stars, explosions of supernovae, and gas transfer in the spiral arms,
so that relatively flat gradients have established. The discovery of the nature of the abundance
step provides a unique opportunity to contribute to the debate on the age of the present 
spiral pattern of the Galaxy, a parameter of major importance for the understanding of the
physics of the spiral arms. A concept of "transient spiral arms" has been supported by a number of papers 
(e.g. \citealp{Sellwood1984}, \citealp{Sellwood2011}). According to that hypothesis,
which is based on numerical simulations, the lifetime of the arms is only of a few Galactic
rotations.  However, based on similar techniques, but with a larger number of particles, 
\cite{Fujii2010}, concludes that the lifetime of a spiral structure can be of up to 10 Gyr.

The existence of the step in abundance means that the corotation resonance stayed at the same
position for a long period. If the spiral structure were a transient phenomenon, changing
its characteristics and its corotation radius at time intervals of the order of a few galactic
rotations,  the abundance step would not exist. We next discuss a lower limit to the age of the
metallicity step. Supposing that a new spiral structure established itself at some time in the past,
with its corotation radius at its present radius, then  different metallicities started to build
up  on both sides of the resonance. A minimum elapsed time can be estimated by supposing 
that the production rate of metals was negligible on the outer side (of course,
if the production of metals were not small on that side, a larger time would be required to build
up the same 0.3 dex step).  Taking the slope of  Fe enrichment as a function of time presented 
by  \cite{Chiappini1997} in their Figure 1, applied to the inner side, in the extreme case in
which the enrichment is unimportant in the outer side, to build a 0.3 dex difference requires
at least 3 Gyr.

This number, which is a lower limit, is in contradiction with the concept of transient spiral
arms, in the shorter pattern life-time version of it. A rough estimation of 3 Gyr as the time
elapsed since the last major change in the spiral structure of the Galaxy was already proposed
by \cite{Amores2009} based on the time required to build up the slope of the metallicity gradient
 of the youngest planetary nebulae.

\section{Conclusions}

The abrupt step in the metallicity gradient at about 8.5 kpc (for $R_0$= 7.5 kpc, or 9.5 kpc 
for $R_0$=8.5 kpc), first reported by \cite{Twarog1997} based on analysis of Fe abundance in
a sample of  Clusters, is confirmed by the additional data collected since 1997, and in particular,
by the high-resolution spectroscopy data. A similar conclusion was already reached by \citet{Carney2005}.
The step is also present in the sample of Cepheids that we analyzed, not only for Fe abundance, but
for the $\alpha$-elements as well. 

 The position of this step coincides with the corotation radius of the main spiral structure and is
 easily explained by the presence of this resonance (the possibility of the existence of a second pattern speed
 was also discussed). As shown by \cite{Amores2009} there is a ring-shaped void of gas at this radius, 
 which was predicted by hydrodynamic simulations (e.g. \citealp{L'epine2001}). This void is caused by the 
 pumping out effect of the spiral structure that produces an inward flow of the gas in the inner regions 
 of the Galaxy and an outward flow  in the outer regions. Therefore, the Cassini-like gap is a barrier
 that isolates the gas of one side from the other, so that the chemical evolution of the two sides 
 are independent. One must remember that chemical evolution models are models for the gas; stars are not
 usually regarded as a mean of transporting metallicity from one Galactic radius to the other. 
 The corotation barrier only acts on the gas, not on the stars; many stars have orbits with strong
 deviations from the circularity, and there is no impediment for them to cross the resonance;
 what is more, stars on circular orbits themselves are scattered across the corotation radius 
\citep{Sellwood2002,L'epine2003}, and thus naturally cross the chemical barrier.

 The counts of  Open Clusters suggest that in the Galactic radius  range 6.2 to 12 kpc, that is, 
 between the ILR and the OLR of the 4-arms mode, but excluding a narrow gap at corotation, the 
 star-formation rate is larger than outside this region. This is not an unexpected result, since
 our Galaxy has an important 4-arms mode, which is only permitted in this region.
 
 Close to corotation, on the inner side of it,  one can find  Open Clusters with differences as 
 large as 0.4 dex in Fe abundance at a same Galactic radius, which is much larger than the
 errors of measurements. The large dispersion of Fe abundance in this region can be attributed to
 the fact that the stars have been originated from different Galactic radii or different star-formation
 regions, through either non-circularity ("blurring") or angular momentum exchange of circular
 orbits ("churning"), making the variety of birthplaces a more important cause of dispersion of metallicity 
 than variations in age, except for the very old clusters.
 
 A fine structure study of the metallicity distribution in populations of clusters and Cepheids
 situated in  a Galactic radius range of width 1-2 kpc, in any  of the two sides of corotation, 
 shows that the pattern of this distribution is not smooth but presents  a number of peaks, shown 
 by horizontal lines in Figure 5. Our interpretation is that the star formation process
 is active in a limited number of regions of the Galactic disk; each active region has its own 
 well defined  metallicity, which is shared by the stars born there. Due to their initial velocities 
 at the instant of birth, the stars (or clusters) have non-circular orbits and usually present
 variations in their Galactic radius of the order of 2 kpc in periods of time of the order of 100 Myr.
 In the external parts of the Galaxy the amplitudes of the radial variations get larger, enhancing 
 the "blurring" effect, but the time-scales are also larger. The simultaneous presence, at a
 given Galactic radius, of stars that originated from different star-forming regions can be
 explained by the overlapping of orbits. The concept is 
 illustrated by L+5, where we see that orbits of different natures can cross each other. In the 
 same paper, the complexity of the local spiral structure is shown, with the young stars of the
 solar neighborhood distributed in spiral arms with many different orientations.

 In other words, the recent chemical enrichment of the Galactic disk seems to have its origin in a clumpy 
 distribution of dominant star-formation centers.  As the stars move out from the star-formation centers,
  and present variations of their Galactic radius, they form the horizontal concentrations of stars with
 similar metallicity like those seen in Figure 5. Certainly  the star formation centers  belong
 to  the spiral arms.
 
 To further convince the reader that one cannot neglect the spiral structure in the interpretation
 of the chemical gradients, we showed that there is a significant azimuthal metallicity gradient in 
 a broad Galactic radius range around the Sun. With the same purpose, we showed an extragalactic example
 of the existence of a step in the gradient, associated with  overlapping gradients (two different metallicity
 curves at a same radius). Analyses of  metallicity gradients in our Galaxy or in external galaxies
 that are restricted to the determination of a single slope as a function of radius are poor ones. 
  
As a conclusion, we thus stress that several difficulties exist in the
comparison of chemical evolution models with observations. Only
recently, such models predicting the metallicity distribution of
both the stars and gas in a realistic manner have been developed
\cite{Schonrich2009}, by taking into account both the angular
momentum exchange across corotation of minor transient spiral features
and the range of radii swept by stars on non-circular
orbits. However, such recent models are still axisymmetric in
essence, and do not take into account some of the basic features
of our non-axisymmetric Galaxy, like the chemical barrier at the
corotation of the main grand-design spiral pattern, the flow of
gas with inverse direction on both sides of it, and a more correct
expression for the star-formation rate taking into account the rate
at which the spiral arms (where stars are born) are fed with gas.
Consequently they cannot at present reproduce features such as the
metallicity step and the overlapping abundance gradients observed
around the corotation of the main spiral pattern, or such as the
azimuthal metallicity gradient observed in the vicinity of the Sun.
More elaborated chemo-dynamical models taking into account the main
effects of the grand-design spiral pattern of the Milky Way are thus
now required to reproduce such observations.

\subsection*{Acknowledgements}
The thank the anonymous referee for several useful suggestions.
The work was supported  by the S\~ao Paulo State agency FAPESP through the PhD
grant 08/06304-3. PC is is now supported by RoPACS, a Marie Curie Initial Training 
Network funded by the European Commission's Seventh Framework Programme. 
BF acknowledges the support of the AvH foundation.

;\bibliographystyle{mn2e}
\bibliographystyle{plainnat}
\bibliography{lepine-final}

\end{document}